\documentstyle[prl,aps,twocolumn,epsf]{revtex}
\def\break#1{\pagebreak \vspace*{#1}}
\begin{document}
\draft
\title{Distribution of sizes of erased loops for loop-erased random walks}
\author{Deepak Dhar $^1$ and Abhishek Dhar $^2$}
\address{Theoretical Physics Group, Tata Institute of Fundamental Research,
Homi Bhabha Road, Bombay 400005, India.\\
$ ^1$ e-mail : ddhar@theory.tifr.res.in
$ ^2$ e-mail : abhi@theory.tifr.res.in}
\date{\today}

\maketitle
\widetext
\begin{abstract}
We study the distribution of sizes of erased loops for loop-erased 
random walks on regular and fractal lattices. We show that for
arbitrary graphs the 
probability $P(l)$ of generating a loop of perimeter $l$ is
expressible in terms of 
the probability $P_{st}(l)$ of forming a loop of perimeter $l$ when a
bond is added to a random spanning tree on the same graph by the
simple relation
$P(l)=P_{st}(l)/l$. On 
$d$-dimensional hypercubical lattices, $P(l)$ varies as
$l^{-\sigma}$ for large $l$, where $\sigma=1+2/z$ for  $1<d<4$, where
z is the fractal dimension of
the loop-erased walks on the graph. On recursively constructed
fractals with $\tilde{d} < 2$ this relation is
modified to $\sigma=1+2\bar{d}/{(\tilde{d}z)}$, where $\bar{d}$ is the
hausdorff and $\tilde{d}$ is the spectral dimension of the fractal.
\end{abstract}


\narrowtext

The loop-erased random walk (LERW) is a simpler variant of 
the well-known self-avoiding walk (SAW) problem, 
which retains the no self-intersection property of SAWs, but is closer to
the ordinary random walk problem. In this paper, we study the
distribution of sizes of erased loops for LERWs on arbitrary graphs
and relate it to the distribution of sizes of loops formed when a bond is
added to a random spanning tree on the same graph. In the themodynamic
limit this distribution has a power law tail, and we express the
exponent in terms of the fractal dimension of the chemical paths on
spanning trees. We show also how this relation is modified for
deterministic fractals.

In spite of the fact that the LERW model is somewhat more tractable 
analytically than the SAW problem, the number of papers devoted to
this problem has remained rather small. 
The model was defined by Lawler\cite{lawl1}, who 
called it the loop-erased self-avoiding walk (LESAW). This terminology
is somewhat inappropriate and we prefer to use the term LERW in this
paper \cite{futnot}. Lawler showed 
that for space dimensionality $d>4$, the large length 
scale properties of the LERWs are same as those of simple random
walks. Thus if $r_n$ is the end-to-end distance of a $n$ step LERW
and we define the exponent $\nu$ such that $<r_n^2> \sim n^{2\nu}$ then $\nu$
has the value $1/2$ in $d>4$.
For $d<4$, Lawler derived the rigorous bound 
that $\nu \ge$ the flory value $3/(d+2)$ for SAW \cite {lawl2}. From
numerical simulations,  
Guttman and Bursill \cite{Gutt} obtained the values 
$ \nu=0.800 \pm 0.003$ in two dimensions and $\nu =0.616 \pm 0.004$ in three
dimensions. The corresponding values for SAWS are $\nu=0.750$ and
$\nu=0.59 \pm 0.004$ in two and three dimensions respectively.
This shows that LERW and SAW are in different universality classes.
Guttman and Bursill conjectured  $ \nu = 4/5 $ in two dimensions   
(see however \cite{Bradley}). This was proved by Majumdar \cite{maj} 
by relating the
LERW problem to that of random spanning trees and showing that the
fractal dimension of LERWs is the same as that of chemical paths in
\break{0.9in}
random spanning trees. In two dimensions the latter is
known using the 
equivalence of spanning trees to the $q\to 0$ limit of the $q$-state Potts 
model from conformal field theory \cite{conig}. 

Bounds on the expected number of erased steps have been
obtained by Lawler \cite{lawler} and it was found that in $\le 4$
dimensions the fraction of steps remaining unerased $\to 0$ as the
number of steps $N \to \infty$. Duplantier obtained exact exponents for
the behaviour of the probability that $k$ LERWs of length $n$ starting
at neighboring points do not intersect and 
also the winding number distribution of a 
LERW \cite{dupl}. Lawler has also shown \cite{lawl4} that the LERW is
equivalent to the Laplacian random walk model studied by Lyklema and
Evertsz \cite{lyml}. 

Our interest in this paper is the distribution of sizes of erased loops for
the LERW problem. The corresponding question for the totally random
walks is the well-known Polya problem \cite{spitzer}. For SAWs the
problem is also 
the well-studied question of enumeration of polymer rings
\cite{review}.
A similar problem is encountered in the context of self-organised
Eulerian walkers model and has been studied recently by Shcherbakov et al 
\cite{shcer}.
We show that the probability $P(l )$ that an erased loop 
has perimeter $l$, equals $P_{st}(l )/l $ where  $P_{st}(l
)$ is the probability that a 
loop of perimeter $l $ is formed when a bond is added to a random spanning
tree. For large $l $, $P(l )$ and $P_{st}{(l )}$ 
are expected to show
power law behaviours, say $P(l ) \sim l ^{-\sigma}$ and 
$P_{st}(l ) \sim l ^{-\tau}$. Then our result implies that 
\begin{equation}
\sigma=\tau +1
\end{equation}
We give scaling 
arguments to derive the exponent $\sigma$ in terms of the
fractal dimension $z$ of chemical paths on random spanning trees. 
For deterministic fractals, this expression is modified, and involves
also the ratio of Hausdorff and spectral dimensions of the fractal.
As a simple illustrative example we consider the Sierpinski Gasket,
and calculate the exponents $z$ and $\tau$ directly from first
principles.  

Consider a $N$ step simple random walk
$\omega=
[{(\omega(0),\omega(1),... \omega(N))}]$, where $\omega(k)$ is the position
of the random walker on the lattice after $k$ steps. Let $j$ be the
smallest value such that $\omega(j)=\omega(i)$ for $0 \le i <j \le N$.
We then obtain a new walk, $\bar
{\omega}=[(\omega(0),\omega(1),..\omega(i),\omega(j+1),...\omega(N))]$,
by deleting all steps between $i$ and $j$. This process, corresponding
to removing loops from $\omega$ in chronological order, is repeated
till a $j$ can no longer be found. The resulting walk is a LERW of
length $n \le N$.

Consider a LERW, on a $d$-dimensional hypercubical $L^d$ torus,
formed from a $N$ step simple random walk by erasing loops. We 
define $P(l ,N,L)$ to be the probability that the $(N+1)$th step
results in erasing a loop of perimeter $l $. 
Let 
\begin{equation}
P(l )=\lim_{L \to {\infty}} \lim_{N \to {\infty}} P(l ,N,L)
\end{equation}

Consider the random walk starting at $O$. After $N$ steps of the walk,
we consider the directed tree formed using last exit bonds from all the sites
visited by the walk, except the endpoint of the walk. This is called
the last exit tree $T_N$ after $N$ steps. For $N>>L^d$, all sites of
the lattice are visited at least once, and $T_N$ is a spanning tree.
It was proved by Broder \cite{brod} that in the steady state all such
spanning trees occur with equal probability. The LERW after $N$ steps
is just the 
directed path from $O$ to the endpoint of the walk along $T_N$.

Now, consider a particular loop ${\cal{L}}$ of $l $ directed bonds
$b_1,b_2...b_l $ (see Fig.1). Let $P({\cal{L}},b_j)$ be the
probability that the $(N+1)$th 
step of the walk will result in formation of the loop ${\cal{L}}$ in the LERW
problem with the $(N+1)$th step being along the bond $b_j$. This
occurs if and only if

(i) the $(N+1)$th step forms the loop ${\cal{L}}$ on $T_N$, with $b_j$
as the last step,

(ii) there is a directed path in $T_N$ from $O$ to the head site of
$b_j$, which does not include any bonds in ${\cal{L}}$.

Let $P_{st}({\cal{L}},b_j)$ denote the probability that (i)
occurs. This probability is easy to compute using the break-collapse
method \cite{fortuin} collapsing the loop $\cal{L}$ to a single point. Hence
it is easy to to see that $P_{st}({\cal{L}},b_j)$ is the same
for all $j$ from $1$ to ${l }$. Thus 
\begin{equation}
P_{st}({\cal{L}},b_j)=P_{st}({\cal{L}})/l ,
\end{equation}
where $P_{st}({\cal{L}})$ denotes the probability that loop
${\cal{L}}$ is formed on $T_N$, whatever the position of the last
step.
To calculate $P({\cal{L}},b_j)$, we have to multiply
$P_{st}({\cal{L}},b_j)$ by the conditional probability
$P(O|{\cal{L}},b_j)$ that (ii) occurs given that loop
${\cal{L}}$ is formed {\it on the spanning tree} with $b_j$ as the last
bond. Thus we have
\begin{equation}
P({\cal{L}},b_j)=P_{st}({\cal{L}},b_j).P(O|{\cal{L}},b_j)
\end{equation}  
As for any spanning tree $T_N$ with end point of walk on ${\cal{L}}$,
the directed path from $O$ must lead 
to one of the sites in the loop ${\cal{L}}$, we must have 
\begin{equation}
\sum_{j=1}^\ell  P(O|{\cal{L}},b_j)=1 ~{\rm{.}}
\end{equation}
Summing  Eq.(4) over $j$ from $1$ to $l$,and using Eq.(5)  we get 
\begin{equation}
\sum_{j=1}^\ell  P({\cal{L}},b_j)=P_{st}({\cal{L}})/\ell .
\end{equation}
Finally, we sum over different shapes and positions of the loop
${\cal{L}}$ having the same perimeter $l $, to get 
\begin{equation}
P(l )=P_{st}{(l )}/l 
\end{equation}

In deriving this result
we have used the fact that $N \to \infty $ limit is taken before the 
$\L \to \infty $ limit. It seems reasonable that the order of limits
in the definition $(1)$ can be interchanged without affecting the
value $P(l )$. However, our proof uses the spanning tree property, and
hence needs modification if $L \to \infty $ limit is taken before the $N \to
\infty $ limit.  

On a square lattice it is easy to calculate $P_{st}{(l )}$ exactly for
small values of $l $ \cite{manna}. We thus find $P(2)=0.25$,
$P(4)\approx 0.03681$ and  $P(6) \approx 0.01034$. We have done Monte-Carlo
simulations and verified these figures to very high accuracy.
In two dimensions, $\tau = 8/5 $ \cite{manna}, and this implies that
$\sigma = 13/5$ for $d=2$. This is also in good agreement with our
simulations.

For $d\le4$, the number of steps of the random walk of $N$ steps that
are still not erased is a negligible fraction of $N$ \cite{lawler}.
Then the expected number of erased steps per step of the random walk
is
\begin{equation}
\sum_{l =2}^{\infty}{l P(l )}=1 ~~~~\rm{for}~~d \le 4 
\end{equation}
 For $d>4$, there is a finite probability that a bond
of the random walk which is generating the LERW will not be erased at 
any future time. Let this probability be called $P_{\infty}$. Then for $d>4$,
the average length of loop erased walk for a random walk of $N$ steps 
increases as $N P_{\infty}$. Since the average lengh erased is
$N\sum_{l =2}^{\infty}{l P(l )}$, this implies that  
\begin{equation}
\sum_{l =2}^{\infty}{l P(l )}=1-P_{\infty}.
\end{equation}  
 Thus, using our relation between the LERWs and 
spanning trees, we are led to the interesting and paradoxical
result that $P_{\infty}$  
can be thought of as the probability that adding a bond at random to a random
spanning tree will not form a loop of finite perimeter, and this is
nonzero if the 
dimension d of the space in which the spanning tree is embedded is
greater than $4$. This result clearly depends on the fact that
the thermodynamic limit of large system size is taken
{\it before} the limit $l  \to \infty$ in the summation. For finite lattices,
adding a bond to a spanning tree must lead to formation of a loop.

We now present a scaling argument to determine $\sigma$ in terms of
the fractal dimension of the LERW $(z=1/\nu)$.
Let $n(l ,N)=$ no. of loops of length $l $ 
generated when random walk is
of $N$ steps. The typical excursion of a random walk of $N$ steps $R$
varies as $N^{1/2}$.

For $1<d<4$, the linear size of the largest loop generated is expected
to be of the order of $R$. Since $l  \sim R^z$ the perimeter of the largest
loop $\sim N^{z/2}$. For large $N$, $n(l ,N)$ grows as $NP(l )$. From
finite size scaling theory, we expect that for large $l $ and $N$, $n(l ,N)$
satisfies the scaling form
\begin{equation}
n(l ,N) \sim \frac{N}{l ^{\sigma}} f(\frac{l }{N^{z/2}}) \rm{,}
\end{equation}
where $f(x)$ is a scaling function. For the
cumulative distribution [no. of loops of size $\ge l $]
\begin{equation}
c(l ,N) \sim \frac{N}{l ^{\sigma-1}} g(\frac{l }{N^{z/2}}). 
\end{equation}
The scaling function $g(x)$ is assumed to be finite at small $x$
and decay rapidly for large $x$ i.e. for $l $ larger than the cutoff
length $N^{z/2}$.
For $l =kN^{z/2}$, where $k$ is a finite constant of order $1$ we must
have $c(l )$ of order $1$. This gives
\begin{equation}
\sigma=1+\frac{2}{z}
\end{equation}
In one dimension, the scaling argument given above breaks down. On a 
linear chain, the erased loops can only be of size $l =2$. On more 
complicated but linear graphs, such as ladders, we can have loops of
arbitrarily large values of $l $, but $P(l )$ decays exponentially with 
$l $, and the size of the largest loop generated does not scale as $R$, as
assumed in the scaling argument. For $d>4$, the LERW is 
approximately a random walk and $\sigma = d/2$.

Note that the scaling relation Eq.(12) does not involve the 
dimension of space $d$ explicitly, but still is valid only for $d$
less than  
the the upper critical dimension $4$. It is interesting to ask how this 
relation needs to be modified to remain valid for 
noninteger values of $d$. We confine our arguments to 
recursively defined fractals, which are explicitly constructed spaces 
having a noninteger $d$.

The scaling argument above is easily extended to work for deterministic
fractals. In this case, for a random walk of $N$ steps, $R \sim
N^{\tilde{d}/{2 \bar{d}}}$  (for $\tilde{d} \le 2$), where $\tilde{d}$
and $\bar{d}$ are the 
spectral and hausdorff dimensions of the fractal. Repeating the above
argument, we then obtain  
\begin{equation}
\sigma=1+\frac{2 \bar{d}}{z \tilde{d}}~~~~~~{\rm{for}}~\tilde{d} \le 2.
\end{equation}

Calculation of the chemical distance exponent $z$ for spanning trees 
is quite straightforward for simple deterministic
fractals of finite ramification index. Since this calculation has not
appeared so far in the literature we describe it briefly below for
the Sierpinski Gasket(SG).
 The exact renormalization equations for spanning trees on the SG
may be deduced from the general recursion equations for the q-state Potts 
model in the limit $q \to 0$ \cite{Dhar}. However, for our
purpose here, it is more convenient to use the recursion equations written
down by Knezevic  
and Vannimenus (KV)\cite{knez} in the context of studying collapse
transition of  
branched polymers on the SG. Only $3$ of the $6$ graphs studied by KV
have no vacant sites, and thus only these have non-zero weights for
the problem of spanning trees (with no other interactions). 
These correspond to the cases where all the three vertices of
the $r$th order triangle are connected to 
each other using bonds within the tree, two are connected to each other 
and not to the third, and all three are unconnected [Fig.2] Let these
weights be  
called $A^{(r)}$, $B^{(r)}$ and $C^{(r)}$ respectively. By definition
$A^{(r)}$ gives the number of spanning trees on the $r$th order
gasket, $B^{(r)}$ gives the number of two-rooted trees with two
vertices connected and $C^{(r)}$ gives the number of three-rooted
spanning trees with all vertices unconnected. From KV, or
directly, the recursion equations for $A$,$B$ and $C$ are easily
written down
\begin{eqnarray}
A^{(r+1)}&=&6 A^{(r)^2} B^{(r)} \nonumber \\    
B^{(r+1)}&=&7 A^{(r)} B^{(r)^2} + A^{(r)^2} C^{(r)}  \nonumber \\
C^{(r+1)}&=&12 A^{(r)} B^{(r)} C^{(r)}+14 B^{(r)^3} \rm{.} 
\end{eqnarray}
The initial values are given by $A^{(1)}=3$,$B^{(1)}=1$,$C^{(1)}=1$.
We define a new variable $X^{(r)}=A^{(r)}C^{(r)}/B^{(r)^2}$. It is easy to
see that $X^{(r)}$ 
satisfies the following recursion equation:
\begin{equation}
X^{(r+1)}=\frac{2X^{(r)}+7/3}{49/36+X^{(r)^2}/36+7X^{(r)}/18} 
\end{equation}
 This equation has the fixed point $X^\star =3$. 
Let $l _a$ and $l _b$ be the average lengths of the chemical paths
connecting the lower two vertices of the $r$th order generating
functions $A$ and $B$ respectively. To find the recursions for $l _a$
and $l _b$ consider, for example, the graph of order $(r+1)$ and type
$B$ shown in Fig.3(a). The probability of this graph is
$A^{(r)}B^{{(r)}^2}/B^{(r+1)}=1/(7+X^\star)=1/10$. The length of the
chemical path 
connecting the vertices is $l _a+2l _b$. 
Thus the contribution of this
graph to $l _b^{(r+1)}$ is $(l _a+2l _b)/10$. Summing up over all relevant
graphs we get:
\begin{eqnarray}
l _a^{(r+1)}&=&5l _a/3+2l _b/3  \nonumber \\
l _b^{(r+1)}&=&6l _a/5+l _b  
\end{eqnarray}
We thus obtain $l _a \rm{,} l _b \sim \lambda ^r$, $\lambda =
(20+\sqrt{205})/15$ being the largest eigenvalue in Eq.(16). Thus 
$l  \sim R^z$ where $R$ is the linear size of the gasket and 
$z=\ln \lambda/{\ln 2}$. Putting in the value of $\lambda$ we get
$z=1.1939...$.

We now find the exponent $\tau$. 
We show in Fig.3. how a loop of the order of $R^z$ may be 
obtained by adding a bond so that the one of the lower order $B$ type
graph becomes a $A$ type graph. Supposing there are $R^{\beta}$ positions
where we could have added the bond in order to get the loop. Then the
probability of this event $\sim R^{\beta}/R^{\bar d}$. Hence we obtain:
\begin{equation}
R^{-z(\tau-1)}=R^{(\beta-\bar{d})}.
\end{equation}
Thus we can find $\tau$ if we can determine the exponent $\beta$,
which gives the fractal dimension of the boundary between the two
constituting sub-trees of 
the $B$ type graph. We note that $R^ {\beta}$  times
$B^{(r)}$ gives the number of ways of getting a $A$ graph by
addition of a bond to a $B$ graph with the added bond labelled. 
But for every resulting $A$ graph
the labelled bond could be anywhere on the backbone (path joining the
three corner vertices) of length of order $R^z$. This gives us
\begin{equation}
B^{(r)}R^\beta =A^{(r)}R^z.
\end{equation} 
Now we note that the resistance between two points on a lattice
with unit resistances on all bonds is given by the ratio of number of
two-rooted spanning trees, with roots at the two given points, to the
number of single-rooted spanning trees. 
It follows then that the ratio $B/A$ gives the 
resistance between the corner points of a triangle, which scales as
$R^{\alpha}$ and so is related to the spectral
dimension of the lattice. It can be shown easily that 
$\alpha=2\bar{d}/{\tilde{d}}-\bar{d}$. Thus from Eq.(18) we get
\begin{equation}
\beta=z-\alpha=z-2\bar{d}/{\tilde{d}}+\bar{d}
\end{equation}
Using this in Eq.(17) we get:
\begin{equation}
\tau=2\bar{d}/{z\tilde{d}}
\end{equation}
From Eq.(7) the LERW exponent $\sigma=\tau+1$ and we verify the result
Eq.(13) obtained by simple scaling arguments.

We thank S. N. Majumdar and M. Barma for critically reading the
manuscript.

\centerline{\bf Figure Captions}

\begin{figure}
\caption{\label{fig1}
A spanning tree $T_N$ with the endpoint of walk at $X$. 
The loop ${\cal{L}}$ (shown in bold) of $10$ directed bonds
is formed if the bond $b_8$, denoted by the white arrow, is added
to $T_N$. In this case there is a directed path from the origin
$O$ to the head of $b_8$.}
\end{figure}

\begin{figure}
\caption{\label{fig2}
 Diagrams representing the generating functions for spanning trees on
the Sierpinski gasket.}
\end{figure}

\begin{figure}
\caption{\label{fig3}
Formation of a loop of perimeter of the order of $R^z$ on
addition of a bond.}
\end{figure}


\begin{thebibliography}{999}
\bibitem{futnot} Firstly, as loop-erased implies loopless, the
qualifier `self-avoiding' is redundant. More importantly, the LERW
problem is related to models such as spanning trees, resistor
networks, which are more directly related to random walks, and not to
self-avoiding walks which belong to a different universality class.
\end{thebibliography}
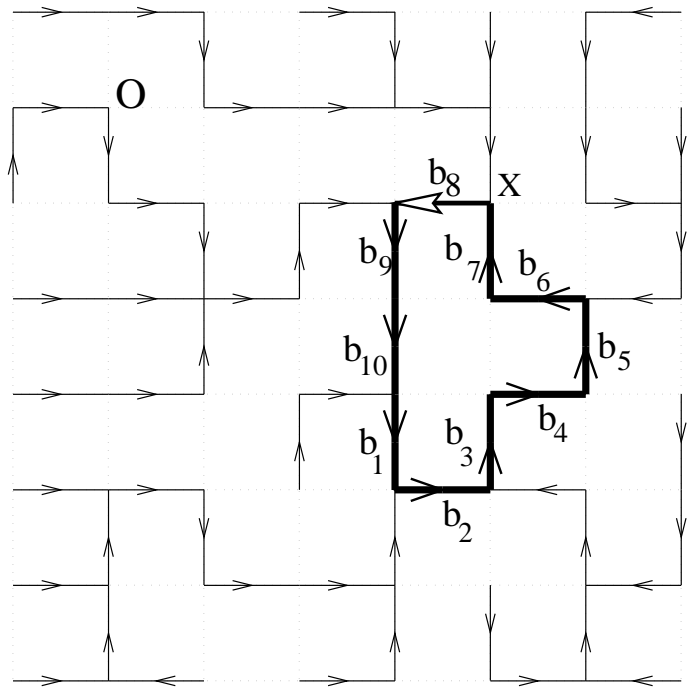
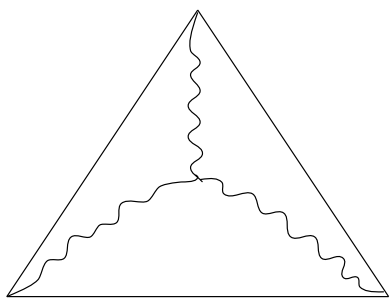
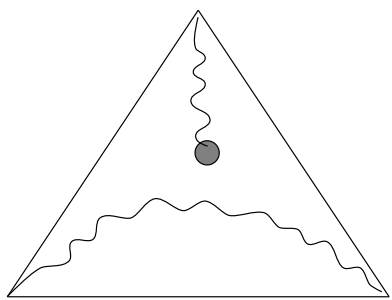
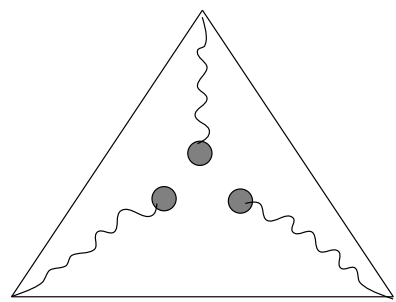
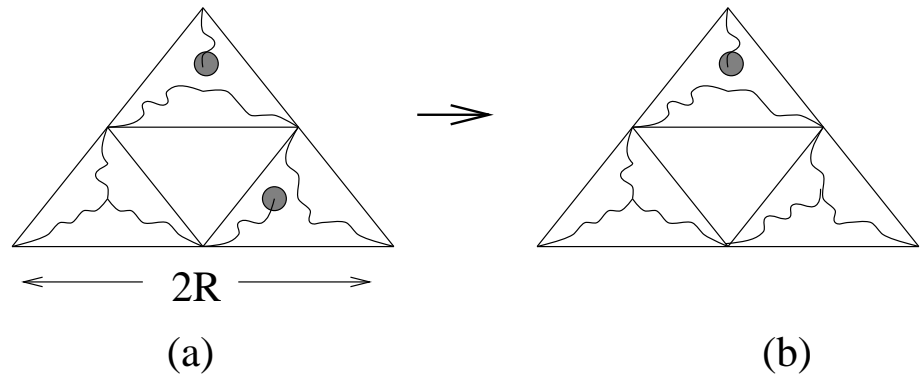
\end{document}